\documentclass[a4paper,UKenglish, cleveref, authorcolumns]{lipics-v2021}

\pdfoutput=1 
\hideLIPIcs  
\title{Investigating Crossing Perception in 3D Graph Visualisation}

\author{Ying Zhang}{University of Konstanz, Germany}{ying.zhang@uni-konstanz.de}{https://orcid.org/0000-0003-4071-0923}{}
\author{Niklas Gröne}{University of Konstanz, Germany}{niklas.groene@uni-konstanz.de}{}{}
\author{Karsten Klein}{University of Konstanz, Germany}{karsten.klein@uni-konstanz.de}{https://orcid.org/0000-0002-8345-5806}{}
\author{Giuseppe Liotta}{University of Perugia, Italy}{giuseppe.liotta@unipg.it}{https://orcid.org/0000-0002-2886-9694}{}
\author{Falk Schreiber}{University of Konstanz, Germany \and Monash University, Australia}{falk.schreiber@uni-konstanz.de}{https://orcid.org/0000-0002-9307-3254}{}

\authorrunning{Y. Zhang, N. Gröne, K. Klein, G. Liotta, F. Schreiber}

\Copyright{Ying Zhang and Niklas Gröne and Karsten Klein and Giuseppe Liotta and Falk Schreiber}



\relatedversion{} %optional, e.g. full version hosted on arXiv, HAL, or other respository/website

\acknowledgements{We acknowledge funding by DFG under Germany’s Excellence Strategy – EXC 2117 – 422037984, and DFG project ID 251654672 – TRR 161; we also acknowledge funding by MUR Project no. 2022TS4Y3N –1405 EXPAND}%optional

\nolinenumbers %uncomment to disable line numbering

\EventNoEds{2}
\EventLongTitle{33nd International Symposium on Graph Drawing and Network Visualization (GD 2025)}
\EventShortTitle{GD 2025}
\EventAcronym{GD}
\EventYear{2025}
\EventDate{December 24--27, 2016}
\EventLocation{Norrköping, Sweden}
\EventLogo{}
\SeriesVolume{42}
\ArticleNo{23}

\newcommand{\pecclong}{projective edge crossing configurations}
\newcommand{\peccshort}{PECC}

\begin{document}
\maketitle

\begin{abstract}
Human perception of graph drawings is influenced by a variety of impact factors for which quality measures are used as a proxy indicator. The investigation of those impact factors and their effects is important to evaluate and improve quality measures and drawing algorithms. The number of edge crossings in a 2D graph drawing has long been a main quality measure for drawing evaluation. The use of stereoscopic 3D graph visualisations has gained attraction over the last years, and results from several studies indicate that they can improve analysis efficiency for a range of analysis scenarios. While edge crossings can also occur in 3D, there are edge configurations in space that are not crossings but might be perceived as such from a specific viewpoint. Such configurations create crossings when projected on the corresponding 2D image plane and could impact readability similar to 2D crossings. In 3D drawings, the additional depth aspect and the subsequent impact factors of edge distance and relative edge direction in space might further influence the importance of those configurations for readability. We investigate the impact of such factors in an empirical study and report on findings of difference between major factor categories.

\end{abstract}

\section{Introduction}
Human perception and understanding of graph drawings is influenced by a variety of impact factors, including aspects of the visual encoding and the characteristics of the drawing's layout. It is hard to capture and quantify all relevant factors in detail, and the mechanisms of graph drawing perception are not yet understood well. There is a recent increase in research to investigate those mechanisms, to evaluate existing measures and methods with respect to perception, and to underpin underlying assumptions with empirical evidence. Up to date, a large list of simple-to-compute quality metrics have been proposed to serve as a proxy indicator for a drawing's quality, with sometimes only marginal support through human studies.
The number of edge crossings has been used and investigated for a long time as a major quality measure. This has lead to theoretical advances, for example in the investigation of computational complexity~\cite{chimani_et_al:LIPIcs.GD.2024.33,schaefer2018crossing}, optimisation~\cite{MUTZEL200833}, and the characterisation of graph classes such as for beyond-planarity~\cite{DBLP:journals/csur/DidimoLM19,hong2020beyond}, and practical advances in layout methods and algorithm engineering~\cite{chimani2021star}. Crossings have been further investigated in empirical studies for their impact on readability to gain a deeper understanding on their influence for graph perception. After the seminal initial experiments on the influence of crossings for readability~\cite{Purchase97,purchase1997aesthetic}, further experiments indicated that their influence might depend on graph characteristics such as size and density~\cite{kobourov2014crossings} or layout characteristics such as the crossings' angles~\cite{huang2013establishing,HUANG2014452,huangEffectsCrossingAngles2008}.
The focus on the number of crossings as a major quality measure has shifted based on the evidence created in these studies, the further development of quality measures such as stress or shape~\cite{Eades15,Gansner05}, and the assumption that a more complex combination of measures might better indicate readability~\cite{Ahmed20,Coleman96,huang2013improving,Ryall97,sponemann2014evolutionary,Wang24}. However, the general questions remain how crossings influence the perception of graph drawings and how the interplay of different drawing factors increases or decreases their influence. 

For network visualisations in stereoscopic 3D (S3D) a number of studies found advantages compared to 2D visualisations~\cite{joos2025visual}. The advantages are however not inherent to the use of S3D, and there are dependencies on the data, representation, environments, and tasks, see e.g.~\cite{feyer20232d,greffardClassicalMonoscopic3D2014,Kwon16,Sorger21}. Thus, both 2D and S3D visualisations for visual network analysis should be considered for use depending on the analysis scenario at hand. 
The investigation of quality measures in S3D is however more complex than for the classical 2D visualisations due to the richer design space and impact factors such as viewer perspective, view mode, and depth. On the one hand, established 2D quality measures might have a different importance in S3D, and their interplay might differ from the 2D scenario; on the other hand, due to the additional impact factors, we might require extensions for their application in 3D or even adopt completely different measures. 
While edge crossings can rarely occur in 3D, edge configurations in space that are not crossings might be perceived as such from a specific viewpoint. Such configurations create crossings when projected on the corresponding 2D image plane. Thus, they could impact readability similar to 2D crossings or, due to additional impact factors such as spatial proximity, orientation of the involved edges, and eye depth focus, have a different effect or even not be perceived as crossings at all. The different depth cues perceived by human observers might subsequently change the way the graph's structure is interpreted. 
To indicate that most of such edge configurations are not edge crossings in 3D per se, but could be perceived as such, we call them `\emph{\pecclong{}}' in the following, or `\peccshort{}' in short, based on the crossings they create in a perspective projection on an image plane.

Given the degree of freedom for \peccshort{}s in 3D, we want to understand which conditions impact their perception for graph analysis and to which extent. A corresponding exploration should investigate which of the measures and impact factors should be considered most important for visualisation and algorithm design, potentially in relation to specific analysis scenario characteristics. As an initial step in that direction, we investigate the influence of \peccshort{}s on task performance in S3D, where we add the depth distance  of the edges at the projected crossing points to the established criteria of number of crossings and crossing angle (here in the projection).
To avoid confusing our results by the interplay of further major factors, we focus the investigation on these criteria independent of the potential effect of perspective changes, i.e.\ we keep the same perspective for the visualisation.Thus, we are not trying to find out differences between perspectives, but as a more fundamental goal investigate the \peccshort{} impact factors for task-performance-related readability in S3D graph visualisations in line with the following research questions.

\begin{itemize}
    \item \textbf{RQ1:} How do \peccshort{}s impact task-performance and readability of graphs in S3D?
    \item \textbf{RQ2:} Do the number of \peccshort{}s and the projection crossing angle incur similar effects on performance as found in previous research for 2D?
    \item \textbf{RQ3:} Do large depth differences in \peccshort{}s lead to large differences for their impact on task performance?

\end{itemize}

As main contributions, we for the first time discuss the potential impact factors related to \peccshort{} perception, describe our hypotheses based on previous research and our own expertise, and present and discuss the results of the first study that investigates the influence of \peccshort{} on task-performance in a VR environment. 
Our results can also serve as a first step to develop guidelines to improve S3D graph layout computation and evaluation. Note that throughout the paper we use the acronym 3D when talking about configurations in three-dimensional space, and S3D when a human observer is involved with a stereoscopic visualisation device.

\begin{figure}
    \centering
    \includegraphics[width=0.5\linewidth]{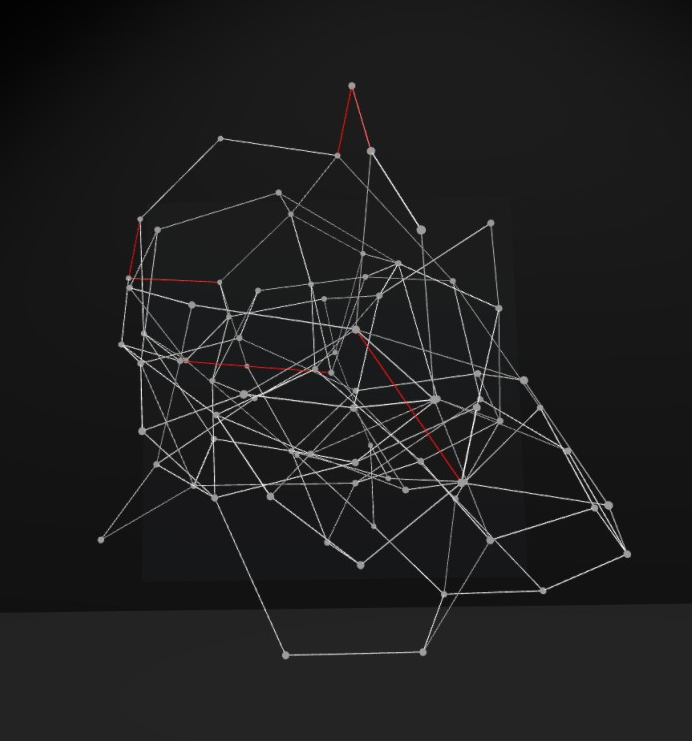}
    \caption{Graph representation for our study in VR.}
    \label{fig:fig1}
\end{figure}

\section{Related Work}

We give a short overview on investigations of graph visualisation perception, crossing effects, and also give examples of general evidence of performance differences between 2D and S3D graph visualisations. These topics partially overlap. For broader discussions of quality criteria, see e.g.\ the overviews in~\cite{Bennett07,burchStateArtEmpirical2021,Dib24}, and for a broader discussions of graph visualisation in immersive environments, see~\cite{joos2025visual}.

\paragraph*{Graph Perception}
Research on perception for graph drawings is still rather scarce despite a recent trend to improve our understanding of the perception mechanisms at play. Most of the related work on perception is concerned with 2D  drawings~\cite{Ballweg18,de2018perception,Grabinger24,MCGRATH1997223,Kypridemou22,Mooney24,Soni18,vanHam08,Wallner20,Ware02}, as the wide use of S3D devices is still a relatively recent phenomenon. 

McGrath et al. observed differences in the interpretation of graph structures based on the differences in the spatial arrangement~\cite{MCGRATH1997223}.
Ware et al.~\cite{Ware02} argued that the investigation of human pattern perception and the consideration of important pattern mechanisms, such as the ones described by Gestalt laws, can help in the evaluation and improvement of quality measures. Going towards a similar direction, Bennett et al.~\cite{Bennett07} provided a discussion on the perceptual base of graph visualisation mechanisms, including the relation to Gestalt principles. Van Ham and Rogowitz observed that people apply rules for perceptual organisation, in particular grouping, when asked to rearrange a graph layout~\cite{vanHam08}. Grabinger et al.~\cite{Grabinger24} found in their study on causal graph perception that the layout had an effect and Gestalt laws allow interpretation of some related aspects.
De Luca et al.~\cite{de2018perception} investigated perceptual recognition of differences in symmetry and found that vertical reflective symmetry was the most dominant. Kypridemou et al.~\cite{kypridemou2020effect} and Soni et al.~\cite{Soni18} investigated the perception of graph characteristics such as density in dependence of the graph layout used, with mixed results on the significance of the layout impact. Mooney et al. investigated the required difference levels for viewers to be able to distinguish two drawings' stress values~\cite{Mooney24}. Bridgeman and Tamassia explored the suitability of measures for similarity evaluation in a user study~\cite{Bridgeman01}. Ballweg et al.~\cite{Ballweg18} explored impact factors for similarity perception of directed acyclic graphs with some evidence on importance of changes at the drawing's borders. Wallner et al.~\cite{Wallner20} also investigate factors for difference perception and found positive impact of added crossings on difference detection.

\paragraph*{Effect of Crossings}
In her investigation of the impact of five quality measures for analysis task performance,  Purchase~\cite{purchase1997aesthetic} found that a lower number of crossings had the most important effect on error rate and completion time for a single small graph and ten different drawings that exhibit a range of the quality measures. Kobourov and Pupyrev~\cite{kobourov2014crossings} showed that the effect might depend on the size and density of the graph. Huang et al.~\cite{huang2013establishing,HUANG2014452,huangEffectsCrossingAngles2008} investigated the impact of the crossing angles, with strong evidence that small angles have a significant negative effect on readability, leading to the research direction of Right-Angle Crossing (RAC) drawings~\cite{DBLP:journals/tcs/DidimoEL11}. Burch et al.~\cite{BurchPartial12} studied a more radical approach to avoid crossing effects by only showing edge stubs. Their findings indicate that this might improve answering time, though with a task-related impact on the error rate.

\paragraph*{3D Graph Visualisation}
 Besides several attempts to use the third dimension for graph visualisation by employing 2D projections of 3D layouts, e.g.~\cite{Dib98,hong2006hierarchical}, there is a long history of investigating graph visualisation in S3D, with several seminal works in the 1990s, such as the investigation of stereo and motion cues by Ware and Franck~\cite{wareEvaluatingStereoMotion1996} and the later follow-up by Ware and Mitchell~\cite{wareReevaluatingStereoMotion2005,wareVisualizingGraphsThree2008}. With increasing availability and quality of stereoscopic devices, further studies investigated differences from 2D and found improvements in S3D, though often in dependence on the graphs, tasks, or representations.
 Kwon et al.~\cite{Kwon16} investigated the use of sophisticated rendering and interaction methods for egocentric visualisations in VR, and found that their technique improved in particular answering time. 
 Belcher et al.~\cite{belcher2003using} compared 2D viewing with AR visualisations using a tangible interface. They observed smaller error rates but higher answering times in AR and confirmed Ware and Franck's conclusions that adding stereo cues provides less advantage than kinetic depth cues.

Greffard et al.~\cite{greffardClassicalMonoscopic3D2014} used S3D projections to study community detection performance and found mostly advantages compared to 2D for more complex structures, while answering time was superior in 2D. Comparing spatial arrangements of multilayer graphs, Feyer et al.~\cite{feyer20232d} reported significant differences in task performance between 2D, 2.5D, and 3D arrangements in a VR environment.
McGuffin et al.~\cite{mcguffin2024path} revisited Ware and Mitchell's study on path tracing 
, 
and also tested a combination of physicalisation and AR. They found that still 3D had lower error rates compared to the 2D condition. In their investigation of the impact of representation fidelity, 
 
Bacim et al.~\cite{Bacim13} found improvement for higher fidelity (with stereo cues and tracking) in a combined score for time and accuracy. In contrast to Ware and Franck, they however did not find significant differences for accuracy alone. Wageningen et al. investigated how 2D quality measures vary across viewpoints sampled on a sphere surrounding 3D graph layouts~\cite{Wageningen2024}.

\section{Methodology}

In order to investigate the effect of \peccshort{}s, we conducted an empirical user study and analysed the collected data with respect to our research questions. In the following, we explain the study design and setup, data collection, tasks, procedure and participants. As our research questions circle around the perception of structures in S3D, a natural choice to present the graph visualisations without further potentially confounding visual distractions is to employ a VR application. For the generated networks and their layouts, spatial arrangement, and further encoding we performed 
small pilot tests to explore reasonable ranges for our settings and parameters with respect to the research questions at hand and the required effects such as sufficient depth perception within the graph representation.

\subsection{Tasks}
The tasks of our study are representatives for different task categories and are standard tasks commonly performed in graph analysis studies, see e.g.~\cite{burchStateArtEmpirical2021}. This allows us to test the impact of \peccshort{} for different analyses and supports comparison with further studies.
We selected three tasks as a compromise to cover a range of categories but keep the study manageable - shortest path finding, common neighbour search, and memorising paths. Note that the latter task deviates from the standard node memorisation task as we want to investigate perceived crossings effects, which requires edges to be involved in the memorised structure.

Thus, the tasks that the participants had to solve were 

\begin{itemize}
    \item (T1) Shortest path - given two highlighted nodes, find a shortest path between them and report its length (in number of edges)
    \item (T2) Common neighbours - given two highlighted nodes, count their common neighbours and report their number
    \item (T3) Memorisation - given a highlighted path, and a highlighted set of edges shown afterwards, decide the overlap and report the number of edges highlighted twice
\end{itemize}

For T1 and T2, we highlighted two nodes, for T3, we first showed a graph with a path highlighted  for $15$ seconds, and after $10$ seconds of a blank screen gap we showed the graph again with the same number of edges highlighted which were either on the path or close-by. See Sections~\ref{sec:setup} and~\ref{sec:stimuli} for further settings and procedure.

\subsection{Potential Impact Factors}
\label{sec:impact}

\paragraph*{\peccshort{} Configuration.}
There is a large number of potential impact factors on graph perception related to \peccshort{}s, for which we needed to make a decision on what to test in our study. These include in particular (controlled ones for the study in bold):
\begin{itemize}
    \item \textbf{(F1)} Number: of \peccshort{}s
    \item \textbf{(F2)} Projection angle: of \peccshort{}s, i.e.\ the angle of the crossing created by a \peccshort{} by projection on the image plane. This is the analogue to the edge crossing angle in 2D. 
    \item (F3) Angle: of edges out of the projection plane (e.g.\ directed towards back/front) 
    \item (F4) Distribution: of \peccshort{}s in the task-relevant region, e.g.\ along the shortest path 
    \item (F5) Position: of involved edges relative to the viewer (e.g.\ distance)
    \item \textbf{(F6)} Distance: of the \peccshort{} edges in the graph representation    
    \item (F7) Length: of edges and ratio of perceived crossing segments on each side.
    \item (F8) Order: of the involved edges in depth direction - is the task-relevant edge to the front or the back of the \peccshort{}?
    \item (F9) Complexity: of the perceived graph structure around the \peccshort{}, such as the density or node degrees, locally or globally
    \item (F10) Visual encoding: e.g.\ size/width or colour of nodes and edges
\end{itemize}

As the edges are situated in 3D space there are several ways of measuring distance. We could use shortest distance of points on the edges or distances between the end nodes. As our investigation is concerned with perceived crossings, we instead use the distance between the two points on the edges in space that constitute the perceived intersection point. 

There are relations between several of these impact factors, e.g.\ when we consider the distance and position of \peccshort{} edges. Depending on their position (F5), the perceived length and crossing segment ratio of the edges (F7) might differ. Further, as each of the factors needs the existence of a \peccshort{}, we can assume they do not have a significant effect completely independent of F1. While all factors might be very interesting for an investigation, many of them are not even well investigated in 2D, e.g.\ the distribution of edge crossings along a shortest path (F4), hindering comparison. F5 is covered to some extent by confining the layout in an exocentric viewpoint to a fixed-size region in front of the viewer, as per our perspective-restricting setting.
F9 covers the context in which the task-related structures are perceived, i.e.\ complexity of the neighbourhood around them. We however focus with our investigation on the factors that are directly related to the task-relevant structures, e.g.~\peccshort{}s involving edges of the shortest path. For the context structure we resort to the generic characterisation by the specific generator model and layout method used. 

Too many impact factors as conditions might on the one hand make it difficult to properly conduct the study, and on the other hand decrease interpretability and generalisability of the results. We decided to include the factors that are directly related to established measures known from investigations of 2D visualisation, number of crossings (F1) and crossing angle (F2), and to add a further measure that is specific to the 3D setting. This allows us to investigate the impact of known factors in S3D, while also investigating further characterisations that are specific to the S3D environment. F6 is particularly interesting for our study as it allows to model levels that have either a clear distinction to 2D (large distance) or are closer to the 2D case (small distance). Thus, we settled for F1, F2, and F6 for the definition of our study conditions.

\paragraph*{Viewpoint Changes.}
The dynamic change of the viewpoint by the viewer might lead to confounding effects as the structure might be perceived differently or even the controlled measures might change due to the movement. Thus, we decided to make the investigation independent of the potential effect of viewpoint changes by restricting the viewpoints that the participants could take.
It is known that the combination of kinetic depth cues and stereoscopy can improve the graph analysis results significantly~\cite{wareReevaluatingStereoMotion2005,Ware08,belcher2003using}. Thus, we did not completely fix the perspective but restricted it such that the viewer could still slightly change it via head movement, but only in a small area for which we pre-calculated that the \peccshort{} measures for the visible scene would not change significantly, e.g.\ the number of projection crossings stays the same. In order to indicate the borders of this area for our participants without disturbing their graph reading process, we opted for a fade-out mechanism. When moving the head across the borders of the allowed region, the graph representation was faded out progressively. The participants were made aware of this before the study and could experience the mechanism during the training phase. While we thus allowed for very small changes in the edge depth-distance to viewer distance ratio and perceived angle based on head movement, the visible scenes were differing strongly regarding the number and configuration of \peccshort{}s across our study conditions. That way we could make sure we still have the advantage of full depth cues but minimised potential confounding influence by movement.

\subsection{User Study Design}\label{sec:Design}
Given our decision on impact factors to investigate, we have as independent variables IV1: number of \peccshort{}s, IV2: \peccshort{} projection angle, IV3: \peccshort{} edge distances. We calculate these variables task-related: Instead of measuring them across the whole graph, we investigate \peccshort{}s that are involving task-relevant edges, i.e.\ (shortest) path edges for T1 and T3 and edges adjacent to the selected nodes for T2.
We test for each of the independent variables two levels at the extreme ends of their range in order to test clearly separable conditions. For IV1, we distinguish between level \texttt{L} with $1$ or $2$ crossings 
and level \texttt{H} with at least as many \peccshort{}s as the task-relevant edges, i.e.\ on average at least one per edge. For IV2, through the natural limit of projection crossing angles at 90 degrees, we consider the average of occurring angles a good indicator if we look at the two well-separated extreme groups of below $30^\circ{}$ (level \texttt{L}) and above $60^\circ{}$ (level \texttt{H}). Level \texttt{H} thus roughly captures the angle range that was considered least impacting the response time in the original 2D path tracing experiments by Huang et al.~\cite{huangEffectsCrossingAngles2008}.
For IV3 we tested the visual appearance of distances across the possible range between $0$ and the graph representation depth, and set the thresholds as a ratio of this depth at $0.1$ (below is \texttt{L}) and $0.3$ (above is \texttt{H}). For a larger H threshold it will be increasingly hard to find a reasonable number of \peccshort{} instances.

This results in eight conditions, represented as \texttt{XYZ} in the following, where \texttt{X} indicates the level for IV1 (\texttt{H} for high, \texttt{L} for low), \texttt{Y} the level of the average IV2, and \texttt{Z} the level of the average IV3. For example, \texttt{HLL} stands for a layout that has a high number of \peccshort{}s on the task-related edges, while having low angle and distance averages.
We performed a within-subjects experiment, i.e.\ each participant solved the tasks in each of the conditions. 
We counter-balanced task and condition order with a balanced Latin-square (Williams) design to control for order effects at two levels (eight condition sequences × three task sequences, With an estimated average completion time of less than one minute per task, the overall experiment time per participant
was expected to be less than 45 minutes, avoiding fatigue effects. As dependent variables we collected task completion time and accuracy (proportion of correct answers).

\subsection{Hypotheses}

Given previous evidence from research in 2D and experiences with 3D graph visualisations, several of the impact factors in Section~\ref{sec:impact} come with intuitive expectations of their impact on performance. For F1, one could expect that a larger number of \peccshort{}s will make readability more difficult and thus lead to a decrease in performance. Similarly, for F2 small angles might make path following and distinction of edges harder, and thus have a negative effect on performance. 
For others, the impact is not that clear. For F6, a larger distance might diminish the impact but with a relevant edge in the back might have the opposite effect. 

For F3 and F7 an edge that points into the depth direction might distract less than one parallel to the projection plane, but if a large part of it points towards the user in front of the task-relevant edge it might be distracting. Thus, we see a lot of potential for further investigations of \peccshort{} impact in extension of our work.

\noindent
Our hypotheses, given the clear separation of high and low factors values in our conditions, are:
\begin{itemize}
\item \textbf{H1} It will be easier to solve tasks with less \peccshort{}s  (H1.1), larger angles  (H1.2), and larger distance  (H1.3). 
\item \textbf{H2} Condition performance would be dominated by \peccshort{} number, angle, and distance in that order.
    
\end{itemize}

\subsection{Study Setup, Implementation, and Procedure}
\label{sec:setup}

We chose a VR environment for the study, as we could fully control the appearance of the environment and graphs to minimise distraction. The participants remained seated during the study, wearing a head-movement tracking VR HMD (Meta Quest 2).
We implemented a Unity application based on the VR graph analysis framework GAV-VR~\cite{KerleMalcharek2023GAVVR-68749}. 
The application shows graph visualisations centred in front of the participant's viewpoint. Regarding distance and scaling we made sure that all graph visualisations were fully within the field of view, with a reasonable padding to ensure that the full graph was visible at all times and the participants were not required to rotate their head to see the peripheral structure.
Except for small head movements, there is no interaction that allows the viewer to change the visual representation of the graphs, only a point-and-click controller interaction to select the UI elements for task responses.
A small panel permanently showed the current task and allowed the participants to indicate that they are ready to answer the task question by clicking a button. When the button is clicked, the completion time is recorded and a number pad is shown that allows the participant to enter the answer. The participants could also decide to indicate that they could not answer the question by pressing a dedicated button.

The study was conducted in a quiet computer laboratory room that was blocked during the time of our study. First, each participant gave informed consent, filled out a demographics form and went through an introduction session that explained the necessary background regarding graphs and tasks.
During the introduction session, the participants were told to perform tasks as accurately and as quickly as they could, while the pace was controlled by the participant.
Then the HMD was fitted and the participants underwent a tutorial session with trial tasks to get accustomed to the setting and study procedure. Participants could ask questions and go back during the tutorial in case there were any issues with their understanding of the tasks and the procedure.
Then the main study was conducted, after which the participants filled a questionnaire including NASA TLX load, study experience, and their personal judgment of task difficulties and crossing impact.
 Each session took about one hour, and each participant received compensation of 18 € for their time and effort.

\subsection{Participants}

We recruited $28$ participants ($12$ male, $16$ female) at our university, aged 19-36 years (avg. 24.46). Thirteen of them reported prior VR experience, and seven reported previous graph analysis experience. All participants reported normal or corrected-to-normal vision. 

\subsection{Stimuli and Encoding}
\label{sec:stimuli}
We considered two main strategies to achieve a set of stimuli that exhibit the different characteristics required from the conditions in our study setting. To introduce those artificially, as, e.g.\ in the initial crossing angle experiment by Huang~\cite{huang2009graph}, or to search for instances that exhibit the characteristics in a drawing that is created with a standard drawing pipeline.

We wanted to focus on the crossings' effect in the context of what otherwise is a typical drawing that could stem from a simple drawing pipeline, i.e.\ configurations that would naturally also occur in practice. Thus, we created a large number of graphs and corresponding drawings and tested them for the required conditions. Overall, we generated over $10,000$ drawings. We experimented with different graph generators and layout methods, such that we do on the one hand get enough instances but on the other hand do not create only extreme graph visualisations, e.g.\ with mostly regular appearance. 
We finally settled with the Newman-Watts-Strogatz random graph model, which allows to create graphs with a good combination of local clustering and shortest path lengths as often found in practice~\cite{NEWMAN1999341}. 
While we expect that there might we differences when investigating other graph classes, we do not expect completely different results within the conditions set in our study, rather task-related variations. 
For the layout we decided to use a standard spring-model layout, as it on the one hand created a sufficient range of instances for our different conditions, and on the other hand a reasonable readability for the tasks, which was tested by us before the study. In order to test for our study conditions, we computed the related measures from a fixed viewpoint on the graph layout. We then applied rotations to the graph layouts to find instances of our conditions for each specific combination of graph instance and layout.  It was surprisingly hard to create reasonable numbers of instances for some of the conditions even with those large numbers (around a $1000$ graphs and over $10,000$ rotations). In particular, for task T1 and T2 condition \texttt{HHH} occurred very often while conditions \texttt{HLX} were rare. For common neighbours, conditions \texttt{LXH} were rarest as all adjacent edges of two nodes would need maximally two \peccshort{}s, high distance, and extreme angles at the same time. Number of occurrences were generally much smaller for conditions \texttt{XLX}, i.e.\ low angles, compared to \texttt{XHX}.

There are several further impact factors beyond our controlled variables that might influence the results which we checked to fall within a certain range.  For task T1 we made sure that there is only one shortest path between the selected nodes, such that for correct answers the path found has the calculated configuration of \peccshort{}s, and restricted the path length to a range of $4$ to $6$ edges to avoid floor and ceiling effects, based on pre-testing. Similarly, for task T2 we restricted the degrees of the two selected nodes to be at least $4$ to avoid trivial instances, with the common neighbours falling into the range $0$-$2$. For tasks T1 and T3 we checked the spatial configurations such that there are no paths that mostly move along the viewing direction, which would make it extremely hard to solve the tasks.

We encoded the graphs with what in our experience is a standard representation, using spheres for the nodes and tubes for the edges. We checked several colour encodings for best readability and contrast in our environment, and coloured both nodes and edges white with a large contrast to the black background room colour (dark grey floor for orientation), see Figure~\ref{fig:fig1}. Highlighted nodes and paths were coloured in red to stick out strongly. Node sizes were selected empirically by testing several sizes for readability and distinction of nodes and edges as well as proper separation of nodes.

\section{Results}
In the following, we present the quantitative results of our study as well as the most relevant qualitative feedback. 

\subsection{Quantitative Findings}
Across all conditions, participants answered correctly on 65.5\%  of the trials, see Table~\ref{tab:accrt}. Per task, accuracy was highest in task T2 (77.7\%) and lowest in task T1 (55.4\%). None of our $28$ participants could solve all 24 trials (3 tasks in 8 conditions) correctly (max. 22 correct answers), with in total two occurrences of indicating not to have found a solution. 
The mean completion time (RT) on correct trials was 16.1 s.

Per task, completion was slowest for task T1 with mean 23.2 s, fastest for task T3 with mean 9.4 s, task T2 mean 15.7 s. There were no outliers in the range of accuracy-completion time combinations that we needed to discard. 

\begin{table}[h]
\caption{\label{tab:accrt}}
\centering
\includegraphics[width=.95\linewidth]{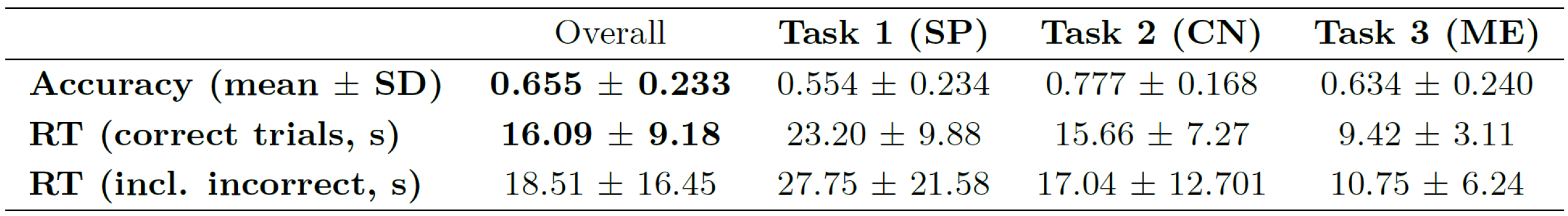}

\end{table}

\begin{figure}[ht!]
    \centering
    \includegraphics[width=0.6\linewidth]{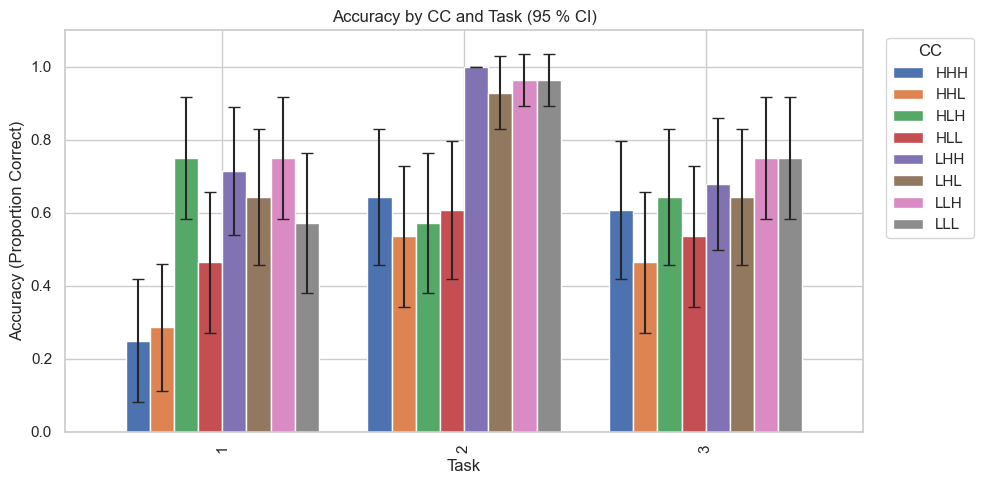}
    \caption{Accuracy by condition and task.}
    \label{fig:performance2}
\end{figure}

Per condition (Table~\ref{tab:CC}), combination \texttt{HHL} yielded lowest mean accuracy ($42.9\%$), with second-slowest mean RT (21s), whereas combination \texttt{LLH} had highest accuracy and shortest completion time (82\%, 12.5~s). Order based on decreasing RT mostly corresponds to order of increasing accuracy with exception of the flipped neighbour pairs \texttt{HHH}/\texttt{HHL} and \texttt{LHH}/\texttt{LLL}. In contrast to the general trend, in particular condition \texttt{HLH} shows large accuracy on a level or even above \texttt{LXX} conditions for tasks T1 and T3, see Figures~\ref{fig:performance2},~\ref{fig:heatmap}. For task difficulty we observe that for task T2 (CN), accuracy is close to 1 for all \texttt{LXX} conditions, and around $0.6$ for the \texttt{HXX} conditions, see Figure~\ref{fig:heatmap}. This indicates that common neighbour counting was pretty easy whenever there was a low number of \peccshort{}. From a binomial generalised linear model (GLM) analysis we derive an odds ratio of $2.59$ for higher accuracy of task T2 compared to T1 ($p=0.001$) and no significant difference between T1 and T3.

\begin{table}[h!]
\centering
\caption{Accuracy and mean RT across all tasks per condition, sorted by RT. Expected impact of count, angle, distance indicated based on hypothesis H1. Highest and lowest values in bold.\label{tab:CC}}
 \includegraphics[width=.9\linewidth]{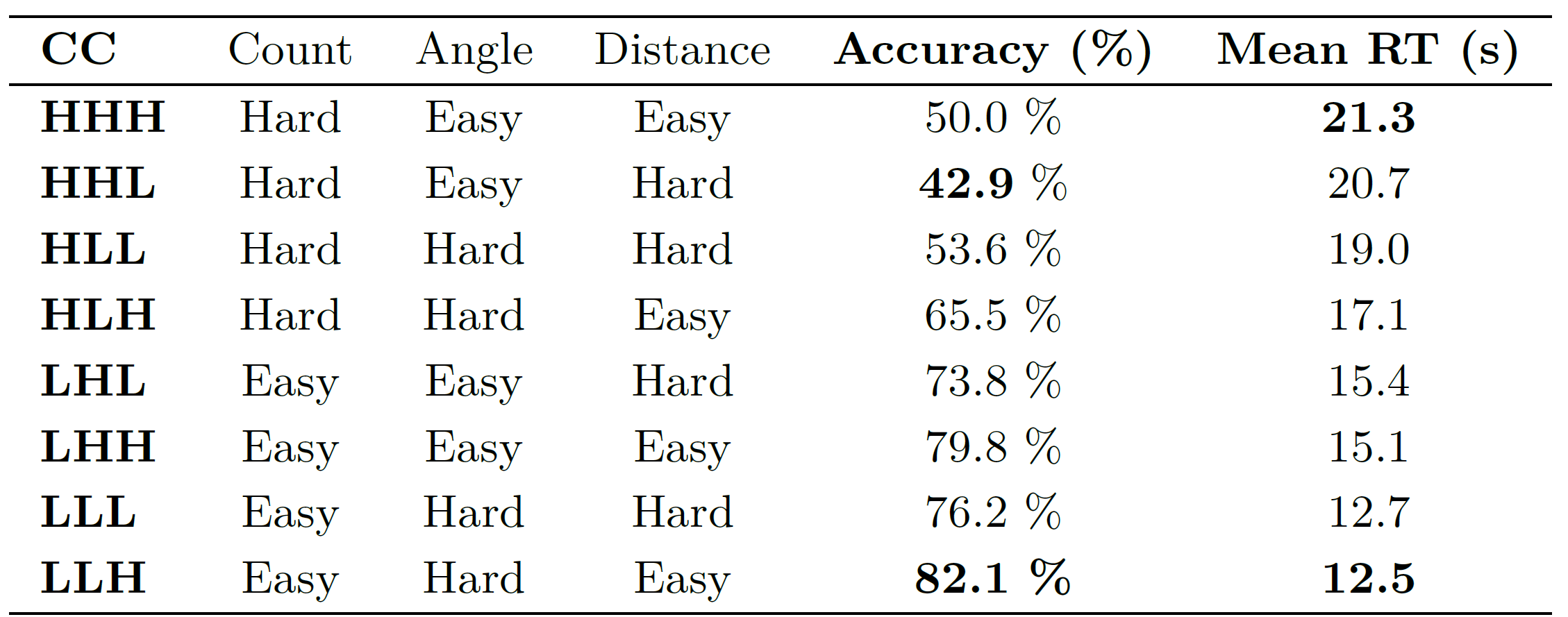}
\end{table}

\textbf{\texttt{HXX} vs \texttt{LXX}} (H1.1): From the mean values, we can see a pattern that \texttt{HXX} combinations have clearly lower accuracy (up to $65.5\%$, which is also the mean) than \texttt{LXX} combinations (at least $73.8\%$), whereas they also have larger completion times. On average, \texttt{LXX} conditions have around $25\%$ higher accuracy than \texttt{HXX} conditions. Similarly, for completion time we have \texttt{LXX} values up to $15.4$ s and \texttt{HXX} values of at least $17.1$ s, with the average for \texttt{LXX} conditions being around $29\%$ lower.
We performed a GLM analysis which supports significance of this observation for accuracy ($p<0.0001$).  To analyse the impact of factor levels per task we performed
a paired Wilcoxon signed-rank test. For task T1 we have significant ($p = 0.004$) higher accuracy for \texttt{LXX} with large effect size ($0.559$, means $0.438$ and  $0.670$). 
For task T2 we have significant ($p = 4.156e-05$) higher accuracy for \texttt{LXX} with large effect size ($0.852$, means $0.589$ and  $0.964$). 
For task T3, we have slightly smaller support but still significant ($p = 0.01902$) higher accuracy for \texttt{LXX} with large effect size ($0.656$, means $0.562$ and  $0.705$). A log-RT mixed model analysis echoes the advantage on speed, showing that the \textbf{\texttt{LXX}} condition reduces completion time ($p < 0.001$). \textbf{Overall, we have significant support for the conclusion that \textbf{\texttt{LXX}} conditions have increased accuracy and reduced completion time across tasks compared to \texttt{HXX}.}

\textbf{\texttt{XHX} vs \texttt{XLX}} (H1.2): However, for the angle impact factor we consistently have the combinations \texttt{XHX} within the two number groups (\texttt{HXX}/\texttt{LXX}) having longer completion time and lower accuracy than their \texttt{XLX} counterparts. GLM analysis supports weak significance ($p = 0.049$) for a positive impact of low angles on accuracy. Split by tasks, we performed a paired Wilcoxon signed-rank test, and the effect is only significant for task T1 ($p = 0.028$) with large effect size ($0.656$, means $0.473$ and  $0.634$). We can see that in particular the \texttt{HHX} combinations have much lower performance than their \texttt{HLX} counterparts, see Figure~\ref{fig:performance2} (left). For the completion time, we have significant low angle speed-ups for tasks T1 ($p = 0.004$) and T2 ($p = 0.021$). \textbf{Overall, we have (weak) significant support for the conclusion that \texttt{XLX} conditions have increased accuracy for task T1 compared to \texttt{XHX}, and partial support for reduced completion time.}

\textbf{\texttt{XXH} vs \texttt{XXL}} (H1.3): For the depth factor we see the expected trend for accuracy--higher values for larger distance--and for completion time close values between the corresponding factor level pairs. Our GLM analysis confirms that the observed accuracy difference is significant ($p = 0.001$). Again, we performed a Wilcoxon test per task and found significant ($p = 0.004$) effects (size $0.764$, means $0.616$ and $0.491$) only for T1.
For completion time, the difference is not significant ($p = 0.056$), and per task we can observe a slow-down for task T1 and slight speed-ups for T2 and T3 with low depth, but all under the significance threshold. \textbf{Overall, we have significant support for the conclusion that \texttt{XXH} conditions increase accuracy for task T1,
%compared to \texttt{XXL}, 
but no support for improvements in completion time.}

\begin{figure}[ht!]
    \centering
    \includegraphics[width=0.6\linewidth]{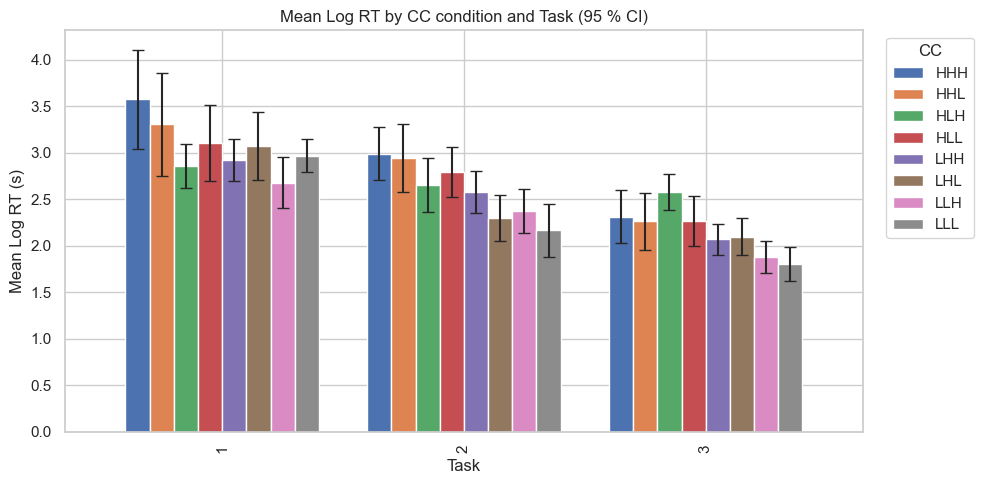}
    \caption{Mean log completion time by condition and task.}
    \label{fig:performance3}
\end{figure}

In summary, we have for accuracy significant differences between high and low \peccshort{} numbers (all tasks), only for task T1 between high and low angles and between high and low distances. For completion time (see Table~\ref{tab:ctime} and Figure~\ref{fig:performance3}) significantly shorter times for low number combinations (\texttt{LXX}) across all tasks, significantly shorter times for low angle combinations (\texttt{XLX}) for tasks T1 and T2, and no significant differences for distance levels. Figure~\ref{fig:Forest} shows the odds ratios for our GLM and LMM analysis.

\begin{figure}[ht!]
    \centering
    \includegraphics[width=0.6\linewidth]{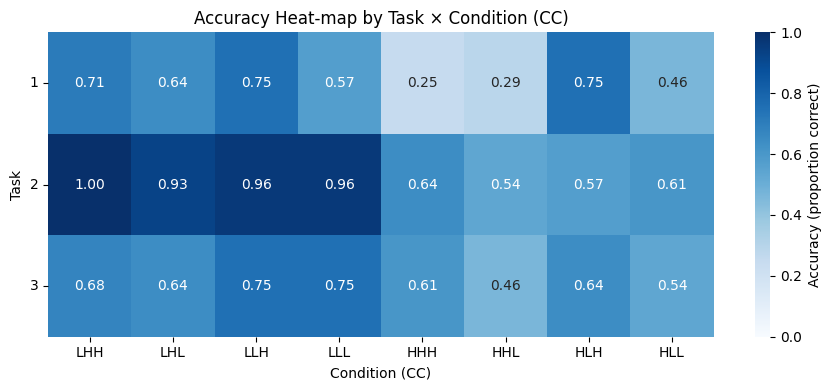}
    \caption{Heatmap for accuracy by task and condition. Clearly visible are the separation of \texttt{LXX} (left) and \texttt{HXX} (right), the high accuracy for T2 and the outlier \texttt{HLH}.}
    \label{fig:heatmap}
\end{figure}

\begin{table}[h!]
    \centering
    \includegraphics[width=0.75\linewidth]{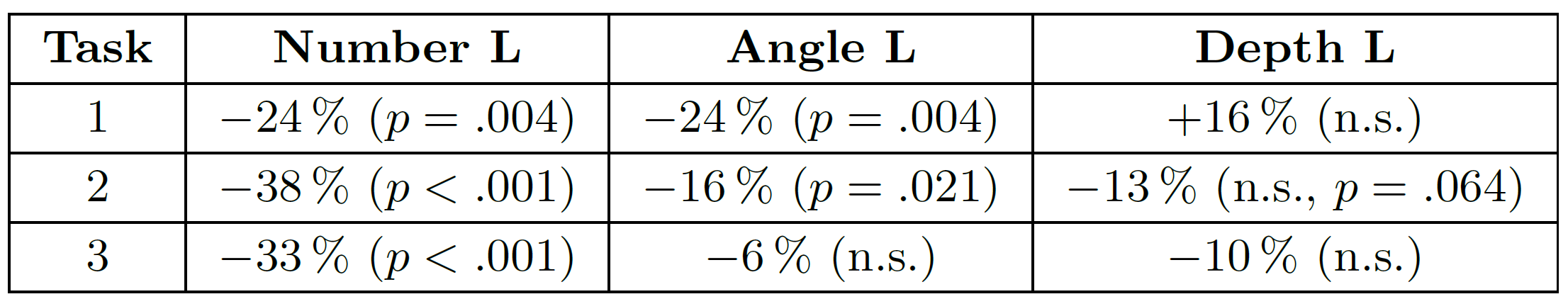}
    \caption{Linear mixed model (LMM) result summary for impact of low factor levels on completion time (log-RT), n.s.= not significant.}
    \label{tab:ctime}
\end{table}

\subsection{Qualitative Feedback}

Participants rated the difficulty of the three tasks on a Likert scale from 1 (easy) to 10 (hard). Task T1 received the highest average ($6.79$, SD$1.68$) with T2 ($4.25$, SD$2.49$) and T3 ($4.86$, SD$2.07$) considered to be much easier, in line with the actual accuracy results obtained.

\begin{figure}[h]
    \centering
    \includegraphics[width=0.45\linewidth]{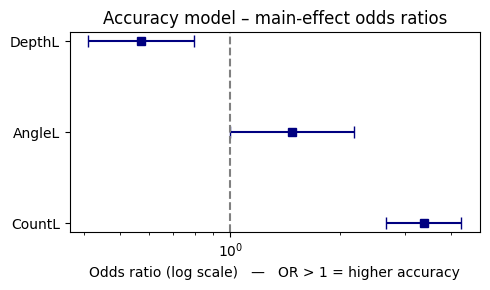}
    \includegraphics[width=0.45\linewidth]{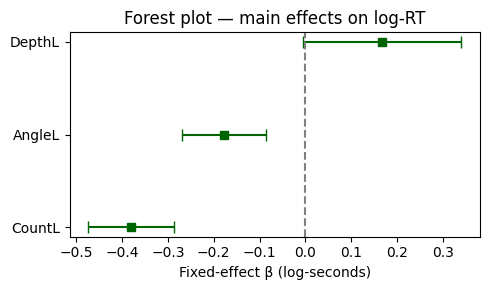}
    \caption{Odds ratios for GLM and LMM analysis of lower factor values; horizontal lines are 95$\%$ CIs. Predictors whose CIs cross OR~=~1 are not significant at $\alpha$ = .05.}
    \label{fig:Forest}
\end{figure}

\section{Discussion}
\subsection{Interpretation of Results}

With respect to RQ1, impact of \peccshort{} on task performance, we could see significant impact on both accuracy and completion time based on differences between pairs of conditions, but also between low and high levels of the number of \peccshort{}. For RQ2, similar effects as in 2D, we thus could observe similar effects for the number, but in contrast to 2D results no similar result for angles. For RQ3, depth impact, we established significant differences between high and low levels of \peccshort{} depth.
Thus, we can confirm hypotheses H1.1 (numbers) across tasks and H1.3 (depth) for shortest path search, but not H1.2 (angles). Instead, much to our surprise we found effects for shortest path search that contrast the previous results from 2D research. We double-checked our analysis to make sure we did not switch angle factor levels, and could not find obvious further confounding impact factors in our instances either. Thus, we think that a deeper investigation of \peccshort{} impact factors should be a topic of future research to improve layout quality measures and algorithms for S3D. A particular setting to compare from our results is the combination \texttt{HHH}/\texttt{HLH} which created the largest discrepancy. As a consequence, we also only have partial support for hypothesis H2 (order of factor effect), where the number factor has clearly the largest effect. 

We can see support for accuracy differences for task T1 between levels across all factors, but not for the other tasks, which might indicate a major influence of the task-solving strategies -- even though task T3 used a path, there is not necessarily a search involved. Regarding the choice of parameters influencing task difficulty, we could see that for task T2 we are close to a ceiling effect. A question for further investigation is if the restriction of common neighbours or degrees play a major role here and if we indeed just approached the lower difficulty limit.

\subsection{Limitations and Future Work}
%As with most empirical studies, t
There are several design parameters in our study that we needed to fix, and several potential impact factors that we could not include as conditions due to the scope of the study. Our study was performed with an exocentric view, and within a well-defined space scale and distance of a cubic space in front of the viewer. While our findings might apply to different settings, e.g.\ viewpoints in an egocentric perspective, one has to be careful in generalising them outside of our setting. A further obvious but necessary restriction is the selection of graphs and layouts, which were fixed to a specific generator model (Newman-Watts-Strogatz), size (80 nodes), and a spring-model layout (Fruchterman-Reingold). Similar to the discussion of edge crossings in 2D, changes in the graph characteristics or the layout might change the influence of \peccshort{} in 3D strongly and should be investigated in further research. As a major future work, the effect of angle should be investigated on a larger scale to confirm that either there is a general mechanism at play that determines our results, or that they are only specific to our particular study settings. A potential impact factor that we did not control is the spatial orientation of crossing edges relative to each other and their length, i.e. measures related to impact factors F3 and F7. We focused on the distance (F6), which will cover some part of this orientation due to the restriction of depth relative to the overall graph representation depth. The difference in direction relative to the viewing direction might also play a role which however should be investigated in an independent study as a careful control is beyond the boundaries of our study design. Further, for a specific distance, the edges relevant for the task solving might be either the front or back edge of a crossing (F8). 
 
We did however not control if the distractor or the task-related edge is at the front. Investigating the impact of these two cases is a further interesting topic to study. Similarly, while we control the number of \peccshort{}s, we did only consider the distribution in post-hoc analysis, e.g.\ how the \peccshort{}s are spread along a shortest path (F4).

\section{Conclusion}
We investigated the influence of \pecclong{} for task performance in S3D graph visualisations through related factors---their number, the angles, and the depth distance. Our setting used an exocentric view, graphs generated by the Newman-Watts-Strogatz model with $80$ nodes, and a 3D spring-model layout. Our main findings on the one hand confirm to our knowledge for the first time the common-sense hypothesis that the number of perceivable crossings also in S3D impedes the readability, and also that higher depth distance for \peccshort{}s diminishes that negative impact. For angles on the other hand, we could not find enough support to confirm positive effects of larger vs. smaller values.

\bibliography{Crossing_Perception_arXiv}

\end{document}